\documentclass[aps,pra,twocolumn,floatfix,showpacs,amsmath,amsfonts]{revtex4}

\usepackage{graphicx}
\usepackage{bm}
\usepackage{dcolumn}
\usepackage{hyperref}

\begin{document}

\title{Exceptional points in the spectra of atoms in external fields}

\author{Holger Cartarius}
\email{Holger.Cartarius@itp1.uni-stuttgart.de}
\author{J\"org Main}
\author{G\"unter Wunner}
\affiliation{Institut f\"ur Theoretische Physik 1, Universit\"at Stuttgart,
  70550 Stuttgart, Germany}
\date{\today}

\begin{abstract}
We investigate exceptional points, which are branch point singularities of
two resonance eigenstates, in spectra of the hydrogen atom in crossed external
electric and magnetic fields. A procedure to systematically search for
exceptional points is presented, and their existence is proven. The properties
of the branch point singularities are discussed with effective low-dimensional
matrix models, their relation with avoided level crossings is analyzed, and
their influence on dipole matrix elements and the photoionization cross section
is investigated. Furthermore, the rare case of a connection between three
resonances almost forming a triple-degeneracy in the form of a cubic root
branch point is discussed. 
\end{abstract}


\maketitle

\section{Introduction}

A special type of degeneracy in parameter-dependent resonance spectra,
namely ``exceptional points'' \cite{Kato66,Hei99}, has recently attracted
growing attention theoretically
\cite{Hei00,Hei90,Ber98,Cej07,Kla08b,Lee08,Mue08}
as well as experimentally \cite{Phi00,Dem01,Dem03,Dem04,Die07,Ste04}.
Exceptional points can be found in systems which depend on at least two real
valued parameters. They are positions in the parameter space at which usually
two energy eigenvalues pass through a branch point singularity, i.e., the two 
eigenvalues can mathematically be described by two branches of the same
analytic function and the exceptional point represents the branch point
singularity. In this case, also the eigenvectors, or wave functions, pass
through a branch point singularity and, hence, are identical. The occurrence of
a geometric phase \cite{Ber84a} is one of the important consequences of
exceptional points \cite{Hei99}.

Within the framework of the linear Schr\"odinger equation, the existence of
resonances, i.e., decaying unbound states is important for the appearance of
exceptional points because the coalescence of two discrete eigenstates with
identical eigenvectors is not possible in the spectra of Hermitian Hamiltonians
with potentials, which describe bound states. Here, always a set of orthogonal
eigenstates, which never can become identical, exists. For resonances, the
situation is different. Their eigenstates can become identical. One possibility
of obtaining the resonances is the complex rotation method
\cite{Rei82,Ho83,Moi98}, which leads to \emph{non}-Hermitian Hamiltonians
in the spectra of which resonances are uncovered as complex eigenvalues.

Physical systems in which exceptional points can appear have to depend on
at least a two-dimensional parameter space. If exceptional points are to be
observable it must be possible to adjust these parameters in a sufficiently
wide range of the parameter space. Furthermore, the complex energy eigenvalues,
typically the positions (frequencies or energies) and widths of resonances,
must be accessible with a high precision. Examples are discussed, e.g., for
complex atoms in laser fields \cite{Lat95}, a double $\delta$ well
\cite{Kor03}, the scattering of a beam of particles by a double barrier
potential \cite{Her06}, non-Hermitian Bose-Hubbard models \cite{Gra08}, or
models used in nuclear physics \cite{Bre99}.
The resonant behavior of atom waves in optical lattices \cite{Obe96} shows
structures originating from exceptional points \cite{Ber98}, and they can be
found in nonlinear quantum systems. The stationary solutions of the
Gross-Pitaevskii equation describing Bose-Einstein condensates exhibits a
coalescence of two states due to the nonlinearity of the equation, which turns
out to be a branch point singularity of the energy eigenvalues and wave
functions \cite{Car08a,Rap09}. However, the phenomenon of exceptional points
in physics is not restricted to quantum mechanics. Acoustic modes in absorptive
media \cite{Shu00} represent a mechanical system, in which branch point
singularities appear. Manifestations of exceptional points can also be seen
in optical devices \cite{Pan55,Ber94a,Wie08,Kla08a}. The most detailed
experimental analysis of exceptional points has been carried out for the
resonances of microwave cavities \cite{Phi00,Dem03,Die07}, which open the
possibility of studying the properties of the complex resonance frequencies
and the wave functions. In particular, the geometric phase has been demonstrated
experimentally \cite{Dem01,Dem04}. 

Atoms in static external electric and magnetic fields are fundamental physical
systems. As real quantum systems they are accessible both to experimental and
theoretical methods and have, e.g., very successfully been used for comparisons
with semiclassical theories \cite{Fre02,Rao01,Bar03a,Bar03b}. They provide
a rich variety of physical phenomena such as Ericson fluctuations, which were
discovered both in numerical studies \cite{Mai92,Mai94} and experiments
\cite{Sta05}. Recently, the existence of exceptional points was reported and a
method to detect them in an experiment with atoms was proposed \cite{Car07b}.
Atoms in external fields are well suited to study the signatures originating
from exceptional points in quantum spectra. The electric and magnetic field
strengths span the two-dimensional parameter space required to provide the
branch points.

It is the purpose of this paper to show how exceptional points can be
detected systematically in spectra of the hydrogen atom in external fields and
to discuss their properties in detail. Numerically exact resonance energies and
eigenstates are determined by the complex rotation method
\cite{Rei82,Ho83,Moi98}. The permutation behavior of the resonances for closed
parameter space loops is used to unambiguously find and verify the branch point
singularities. Examples for exceptional points found in the system are
presented. The shapes of the loops of the complex eigenvalues representing the
resonances are explained with a two-dimensional matrix model. It is shown that
single dipole matrix elements of two states coalescing at an exceptional point
diverge whereas this behavior does not carry over to the observable
photoionization cross section.

Besides the typical case, where an exceptional point consists of two
resonances forming a square root branch point, higher degeneracies are
possible. In this paper the rare case of a structure with three resonances
almost forming a triple-coalescence in the form of a cubic root branch point
is presented. It is shown how exceptional points are related to avoided
crossings of the energies or widths of the resonances. 

The paper is organized as follows. Sec.\ \ref{sec:ep_introduction} summarizes
the most important characteristics of exceptional points and demonstrates them
with the help of an illustrative example. The Hamiltonian of the crossed-fields
hydrogen atom is introduced in Sec.\ \ref{sec:hydrogen_hamiltonian}, and the
method to obtain the resonance eigenstates is presented. A procedure to find
exceptional points in spectra of atoms in external fields is presented and
applied in Sec.\ \ref{sec:hydrogen_ep}, in which also examples of branch point
singularities are given. A two-dimensional matrix model used to explain the
paths of the eigenvalues for parameter space loops is introduced and applied
in Sec.\ \ref{sec:hydrogen_properties}. The influence of exceptional points on
dipole matrix elements and the photoionization cross section is discussed in
Sec.\ \ref{sec:hydrogen_dip_behavior}. Three resonances almost forming a
triple-coalescence are presented in Sec.\ \ref{sec:hydrogen_properties_three},
and the connection between exceptional points and avoided level crossings is
investigated in Sec.\ \ref{sec:hydrogen_avoided}. Conclusions
are drawn in Sec.\ \ref{sec:conclusion}.

\section{Exceptional points}
\label{sec:ep_introduction}

\subsection{Important properties}
\label{sec:ep_linear}

The typical case of an exceptional point is a position in a two-dimensional
parameter space at which two eigenvalues pass through a branch point
singularity. This behavior has important consequences for the associated
eigenvalues and eigenvectors.

If one encircles the exceptional point in the parameter space, a typical
permutation behavior of the eigenvalues can be observed \cite{Kato66}.  The
two eigenvalues which represent the two branches of one analytic function with
the singularity are interchanged after one circle around the exceptional point,
whereas all further eigenvalues do not undergo a permutation.
Away from the exceptional points the branching eigenvalues are different and
each of them belongs to a distinct eigenvector. Accordingly, these eigenvectors
undergo the same permutation as the eigenvalues, and at the exceptional points
they likewise pass through a branch point singularity \cite{Kato66}, i.e.,
there is only one linearly independent eigenvector for the two degenerate
eigenvalues. In a matrix representation, the eigenvalues and eigenvectors form
a normal block \cite{Gue07,Sey05}, which distinguishes them from ordinary
degeneracies where two eigenvalues belonging to two different analytic
functions have the same value. Alongside the permutation of the eigenvectors
a geometric phase appears. In the most common physical situation of a square
root branch point resulting from a complex symmetric matrix, the geometric
phase becomes manifest in the change in sign of one of the two eigenvectors
\cite{Hei99} and can be written in the form
\begin{equation*}
  [\bm{x}_1, \bm{x}_2] \overset{\text{circle}}{\to} [\bm{x}_2, -\bm{x}_1] \; ,
\end{equation*}
where $\bm{x}_1$ and $\bm{x}_2$ are the eigenvectors permuted during the loop
around the exceptional point.

It should be mentioned that the occurrence of the phenomenon is not restricted
to two dimensions. In higher-dimensional parameter spaces of complex matrices,
an exceptional ``point'' is always an object of co-dimension 2 \cite{Sey05}.
That is, in two dimensions an exceptional point indeed appears as a point,
whereas in three dimensions it is a one-dimensional object, or line, and in a
four-dimensional parameter space it has the form of a two-dimensional
``exceptional surface.''

\subsection{A simple example in a non-Hermitian linear map}
\label{sec:ep_simple_model}

One of the simplest examples in which an exceptional point occurs, and
which helps to understand many effects related to it, is described by the
two-dimensional matrix  \cite{Kato66}
\begin{equation}
  \bm{M}(\kappa) = \left ( \begin{array}{rr}
      1 & \kappa \\
      \kappa & -1
    \end{array} \right )
  \label{eq:ep_most_simple_model}
\end{equation}
with the complex parameter $\kappa$. The two eigenvalues of the matrix
are given by
\begin{equation}
  \lambda_1 = \sqrt{1+\kappa^2} \; , \quad
  \lambda_2 = -\sqrt{1+\kappa^2} \label{eq_most_simple_ev}
\end{equation}
and are obviously two branches of the same analytic function in $\kappa$.
There are two exceptional points in the system, which appear at the complex
conjugate values $\kappa_\pm = \pm \mathrm{i}$ as can easily be seen. The
eigenvectors which belong to the two eigenvalues are
\begin{equation}
  \bm{x}_{1,2}(\kappa) = \left ( \begin{array}{c} 
      -\kappa \\ 1\mp\sqrt{1+\kappa^2} \end{array} \right ) \; .
  \label{eq:ep_most_simple_model_vec}
\end{equation}
They also depend on the parameter $\kappa$ and pass through a branch 
point singularity at the exceptional points $\kappa_\pm = \pm \mathrm{i}$,
where the only linearly independent eigenvector reads $\bm{x}(\pm \mathrm{i}) 
= ( \mp \mathrm{i}, 1 )$.

\begin{figure}[tb]
  \centering
  \includegraphics[width=\columnwidth]{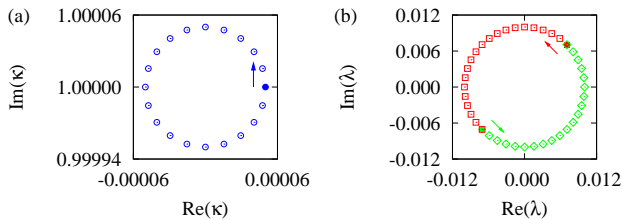}
  \caption{\label{fig:ep_most_simple_model} (Color online) (a) Circle in
    the parameter space $\kappa$ with the exceptional point $\kappa_+ 
    = \mathrm{i}$ as center point for the simple model 
    \eqref{eq:ep_most_simple_model}. 
    (b) Eigenvalues $\lambda_{1,2}$ calculated for the parameter values from
    (a) indicated by red squares and green diamonds, respectively. In this
    case each of the two eigenvalues traverses a semicircle. In (a) and (b)
    the filled symbols represent the first parameter value $\kappa_0$ and
    the corresponding eigenvalues $\lambda_{1,2}(\kappa_0)$, respectively.
    The arrows point in the direction of progression.}
\end{figure}

The branch point singularity leads to a characteristic behavior of the
corresponding eigenvalues under changes of the parameters. If one chooses
a closed loop in the parameter space and calculates the eigenvalues for a set
of parameters on this loop, the permutation of the eigenvalues can be
seen by plotting their paths in the complex energy plane. The situation is
illustrated in Fig.\ \ref{fig:ep_most_simple_model}. 
Here, a circle $\kappa(\varphi) = \mathrm{i} + \varrho
\mathrm{e}^{\mathrm{i} \varphi}$ around the singularity $\kappa_+ = \mathrm{i}$ is
traversed in the parameter space, which is shown in Fig.\ 
\ref{fig:ep_most_simple_model}(a). After one revolution, the first eigenvalue
marked by red squares has traveled to the starting point of the second 
marked by green diamonds, and vice versa. As a consequence, the path of each
eigenvalue is not closed if one traversal of the loop in the parameter space is
performed, but the path is closed if the parameter space loop is traversed
twice. Of course, for the simple model it is also possible to demonstrate the
half-circle structure analytically. If the parameter space curve described
above is applied to the eigenvalues \eqref{eq_most_simple_ev} one obtains, for
$\varrho \ll 2$, the expansion
\begin{align}
  \lambda_{1,2} &= \pm \sqrt{1+(\mathrm{i} + \varrho \mathrm{e}^{\mathrm{i}
      \varphi})^2} = \pm \sqrt{\varrho} \mathrm{e}^{\mathrm{i} \varphi/2}
  \sqrt{2\mathrm{i} + \varrho  \mathrm{e}^{\mathrm{i} \varphi}} \notag \\ 
  &\approx  \pm \sqrt{2 \varrho} \mathrm{e}^{\mathrm{i} \pi/4}
  \mathrm{e}^{\mathrm{i} \varphi/2}
  \label{eq:most_simple_model_expansion}
\end{align}
or
\begin{equation}
  \lambda_1 = \sqrt{2 \varrho} \mathrm{e}^{\mathrm{i} (\pi/4 + \varphi/2)} \; ,
  \quad 
  \lambda_2 = \sqrt{2 \varrho} \mathrm{e}^{\mathrm{i} (5\pi/4 + \varphi/2)} \; ,
  \label{eq:ep_most_simple_value}
\end{equation}
which reproduces the half-circles shown in Fig.\
\ref{fig:ep_most_simple_model} for a full parameter space loop
$\varphi = 0 \dots 2\pi$ correctly.

\section{Hamiltonian and matrix representation}
\label{sec:hydrogen_hamiltonian} 

In this paper the hydrogen atom in static crossed electric and magnetic
fields is investigated from the point of view of exceptional points. The
electric field is assumed to point in the $x$-direction and the magnetic field
is orientated along the $z$-axis. The Hamiltonian without relativistic
corrections and finite nuclear mass effects \cite{Sch93} reads in atomic
Hartree units 
\begin{equation}
  H = \frac{1}{2}\bm{p}^2 - \frac{1}{r} + \frac{1}{2} \gamma L_z
  + \frac{1}{8} \gamma^2 (x^2 + y^2) + f x \; ,
  \label{eq:hydrogen_hamiltonian_atomic}
\end{equation}
where $\bm{p}$ is the kinetic momentum of the electron, $r$ its distance from
the nucleus, and $L_z$ the $z$-component of the angular momentum. 
The Hamiltonian contains the Coulomb potential $\propto 1/r$, the paramagnetic
term $\propto \gamma L_z$, the diamagnetic term $\propto \gamma^2(x^2+y^2)$, and
the potential due to the external electric field $\propto f x$.

For the numerical calculation of the energy eigenvalues in a matrix
representation of the Hamiltonian \eqref{eq:hydrogen_hamiltonian_atomic}
it is advantageous to transform the Schr\"odinger equation into dilated
semiparabolic coordinates,
\begin{equation}
  \mu = \frac{1}{b} \sqrt{r+z}\; , \quad \nu = \frac{1}{b} \sqrt{r-z}\; , 
  \quad \varphi = \arctan\frac{y}{x} 
\end{equation}
with $r = \sqrt{x^2+y^2+z^2}$. For calculating the resonances of the system the
complex rotation method \cite{Rei82,Ho83,Moi98} is used as a powerful
tool \cite{Del91,Mai92,Mai94}. The complex scaling is included via the dilation
parameter $b$, where the replacement
\begin{equation}
  b^2 = |b^2|\mathrm{e}^{\mathrm{i} \theta}
\end{equation}
entails the complex rotation of the position vector
\begin{equation}
  \bm{r} \to \mathrm{e}^{\mathrm{i} \theta} \bm{r} 
  \label{eq:crot_r}
\end{equation}
and leads to the complex scaled and regularized Schr\"odinger equation
\begin{subequations}
  \begin{multline}
    \bigg \{ -2 H_0 + |b|^4 \mathrm{e}^{\mathrm{i} 2 \theta} \gamma \left ( \mu^2 
      + \nu^2 \right )  \mathrm{i}\frac{\partial}{\partial \varphi} 
    - \frac{1}{4} |b|^8 \mathrm{e}^{\mathrm{i} 4 \theta} \gamma^2 \\
    \times \mu^2\nu^2 \left ( \mu^2 + \nu^2 \right ) 
    - 2 |b|^6 \mathrm{e}^{\mathrm{i} 3 \theta} f \mu\nu \left ( \mu^2 + \nu^2
    \right ) \cos \varphi \\
    + 4|b|^2 \mathrm{e}^{\mathrm{i} \theta} 
    + \left ( \mu^2 + \nu^2 \right ) \bigg \} \mathrm{e}^{- \mathrm{i} 2\theta}
    \psi = 2 |b|^4 E \left ( \mu^2 + \nu^2 \right ) \psi
    \label{eq:hydrogen_Schroedinger_complex}
  \end{multline}
  with the term
  \begin{equation}
    H_0 = -\frac{1}{2} (\Delta_\mu + \Delta_\nu) + \frac{1}{2} \left ( \mu^2 
      + \nu^2 \right ) \; ,
  \end{equation}
  where
  \begin{equation}
    \Delta_\varrho = \frac{1}{\varrho} \frac{\partial}{\partial\varrho} \varrho
    \frac{\partial}{\partial\varrho} + \frac{1}{\varrho^2} 
    \frac{\partial^2}{\partial\varphi^2} \;, \qquad \varrho \in \left \{ \mu,
      \nu \right \} \; .
  \end{equation}
\end{subequations}
Due to the harmonic oscillator structure of $H_0$ a well suited complete basis
set is given by the states
\begin{equation}
  | n_\mu, n_\nu, m \rangle = | n_\mu, m \rangle \otimes | n_\nu, m \rangle
  \; , \label{eq:hydrogen_matrix_basis_states}
\end{equation}
where each of $| n_\mu, m \rangle$ and $| n_\nu, m \rangle$ represents an
eigenstate of the two commuting operators 
\begin{subequations}
  \begin{align}
    N &= a_1^+ a_1 + a_2^+ a_2 \; , \\
    L &= \mathrm{i} (a_1 a_2^+ - a_1^+ a_2) = (q_1 p_2 - q_2 p_1)
  \end{align}
\end{subequations}
of the two-dimensional isotropic harmonic oscillator with common eigenvalue
$m$ of $L$. The operators $a_i$ and $a_i^+$ are the familiar ladder operators
of the one-dimensional harmonic oscillator.

The matrix representation of the Schr\"odinger equation
\eqref{eq:hydrogen_Schroedinger_complex} is non-Hermitian and has the form 
\begin{equation}
  \bm{A}(\gamma,f) \Psi = 2 b^4 E \bm{C} \Psi \; ,
  \label{eq:hydrogen_non_Hermitian_matrix}
\end{equation}
where $\bm{A}(\gamma,f)$ is a complex symmetric matrix and $\bm{C}$ is real
symmetric positive definite. Above the ionization threshold, resonances are
uncovered as complex energy eigenvalues $E$, where the real and imaginary parts
represent the positions and the widths $\Gamma = - 2 \mathrm{Im}(E)$,
respectively.

The Hamiltonian has two constants of motion, namely the energy and the
parity with respect to the $(z=0)$-plane. The latter symmetry opens the
possibility of classifying eigenstates by their $z$-parity and to consider
the associated subspaces separately. The examples discussed in this paper are
given for even $z$-parity.

The computation of the eigenvalues was performed using the ARPACK library
\cite{Leh98}. For typical calculations, the number of basis states was on the
order of $10{,}000$ to $12{,}300$. The exact determination of the positions of
exceptional points up to four valid digits often required larger matrices with
up to $17{,}300$ states. If the matrices $\bm{A}$ and $\bm{C}$ are built up
appropriately, they possess a band structure, which was exploited in the 
numerical diagonalizations.

\section{Exceptional points in spectra of the hydrogen atom in external
  fields}
\label{sec:hydrogen_ep}

\subsection{Identification of exceptional points}
\label{sec:hydrogen_search}

The Hamiltonian in the Schr\"odinger equation
\eqref{eq:hydrogen_Schroedinger_complex} is non-Hermitian. Two
parameters, viz.\ the strengths of the external electric and magnetic fields,
are available to influence the positions of the resonances and thus it should
be possible to produce degeneracies of the complex resonance energies. 
Exceptional points do exist in atomic spectra if the fields can be chosen in
such a way that a coalescence of two states occurs. The crossed-fields hydrogen
system fulfills all necessary conditions for the appearance of exceptional
points, however, one has to find them in the spectrum to really prove their
existence. 

A successful procedure to search exceptional points can be found if one
exploits their properties. The permutation of the two eigenvalues involved in
the singularity provides a clear signature which can be used to detect
exceptional points. A good choice for a closed loop is a ``circle'' in the
parameter space of the two field strengths with a radius $\delta < 1$ chosen
relative to a center $(\gamma_0,f_0)$,
\begin{equation}
  \gamma(\varphi) = \gamma_0 (1 + \delta \cos \varphi )\; , \quad
  f(\varphi) = f_0 (1 + \delta \sin \varphi )
  \label{eq:parameter_circle} \; .
\end{equation}
It opens the possibility of scanning a larger area of the parameter space,
namely the complete circular area, at once. A fundamental advantage of the
method is that it allows for automatizing the procedure up to a certain extent.
If the steps on the circle are chosen small enough, the resonances of two
consecutive steps can be assigned to each other unambiguously, and their motion
in the complex energy plane can be traced. Eigenvalues which do not return to
their starting point once the circle in parameter space is closed but are
interchanged with a further resonance are a proof for the existence of an
exceptional point. The exact position of the degeneracy can be determined by
minimizing the distance of the two eigenvalues. After the degeneracy has been
found numerically, a last circle with a small radius (typically $\delta
\approx 10^{-12}$) around the parameter point at which the degeneracy occurs
is used to decide whether or not the branch point singularity structure is
present and the degeneracy found is an exceptional point. 

The geometric phase appearing with exceptional points is accessible through the
eigenvectors representing the resonances in the numerical calculations, and
provides a further possibility of verifying their existence \cite{Car07b}. As
was also shown in Ref.\ \cite{Car07b} the permutation of the complex resonance
energies opens the possibility of detecting exceptional points in experiments
with atoms once the complex energies have been extracted from the
photoionization cross section.

\subsection{Examples}
\label{sec:hydrogen_examples}

With the method described above, exceptional points have been found in spectra
of the hydrogen atom in static external fields \cite{Car07b}.
Table \ref{tab:hydrogen_exceptional_points} lists 17 examples. 
\begin{table}[tb]
  \caption{\label{tab:hydrogen_exceptional_points}Examples for exceptional
    points in spectra of the hydrogen atom in crossed magnetic ($\gamma$)
    and electric ($f$) fields. All values in atomic units. The numbers
    are used as labels to identify the exceptional points.}
  \begin{ruledtabular}
    \begin{tabular}{r|D{.}{.}{7}D{.}{.}{9}D{.}{.}{6}D{.}{.}{8}}
      & \multicolumn{1}{c}{$\gamma$} 
      & \multicolumn{1}{c}{$f$} 
      & \multicolumn{1}{c}{$\mathrm{Re}(E)$} 
      & \multicolumn{1}{c}{$\mathrm{Im}(E)$} \\
      \hline
      1  & 0.002335 & 0.0001177 & -0.01767 & -0.000103 \\
      2  & 0.002575 & 0.000117114 & -0.015067 & -0.0000823 \\
      3  & 0.002752 & 0.0001298 & -0.015714 & -0.00022637 \\
      4  & 0.0030152 & 0.0001231 & -0.01209 & -0.000099 \\
      5  & 0.003045  & 0.0001332 &  -0.015812 & -0.0001896 \\
      6  & 0.0030460 & 0.000127302 & -0.017624 & -0.000087 \\
      7  & 0.0037915 & 0.0001535 & -0.01240 & -0.000164 \\
      8  & 0.004604 & 0.0002177 & -0.022135 & -0.00006878 \\
      9  & 0.004714 & 0.00021529 & -0.01394 & -0.00010 \\ 
      10 & 0.00483 & 0.000213 & -0.01255 & -0.00030 \\
      11 & 0.00529  & 0.0002011 & -0.0150 & -0.000136 \\
      12 & 0.00537 & 0.000214 & -0.01884 & -0.0000679 \\
      13 & 0.005388 & 0.0002619 & -0.02360 & -0.00015 \\
      14 & 0.00572 & 0.000256 & -0.01984 & -0.000258 \\
      15 & 0.00611 & 0.000256 & -0.01593 & -0.00024 \\
      16 & 0.00615 & 0.000265 & -0.0158  & -0.000374 \\
      17 & 0.00776 & 0.000301 & -0.0179 & -0.000756 \\
    \end{tabular}
  \end{ruledtabular}
\end{table}
In the calculations the relative difference $|E_1 - E_2|/|E_1|$ of the two
eigenvalues could be reduced down to $10^{-13}$. However, this is only the
result for a single matrix representation. What is more crucial is the
influence of the complex rotation on the matrix with finite size. Using up to
$17{,}300$ states the convergence of typically three to four valid digits in the
parameters as well as in the energies can be achieved. The convergence was
checked with the stability of the results against changes in the matrix size
and the complex parameter $b$.

\begin{figure}[tb]
  \centering
  \includegraphics[width=\columnwidth]{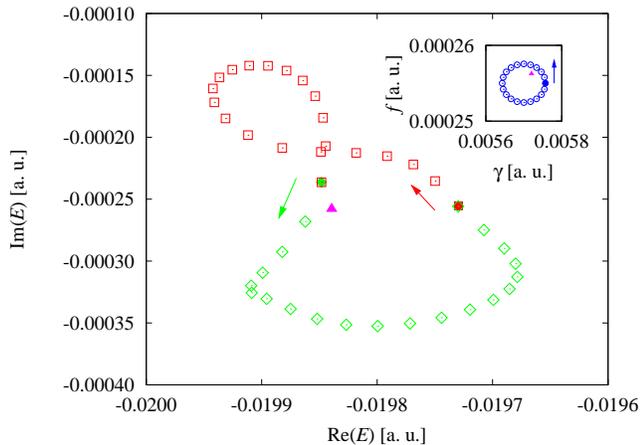}
  \caption{\label{fig:hydrogen_resonances_circle}(Color online) Paths of the
    two eigenvalues (represented by squares and diamonds, respectively) which
    degenerate at the exceptional point labeled 14 in
    Table \ref{tab:hydrogen_exceptional_points} in the complex energy plane
    \cite{Car07b}. Each point of one eigenvalue belongs to a different set of
    parameters. The path in the field strength parameter space is a circle
    defined in Eq.\ \eqref{eq:parameter_circle} with $\delta=0.01$ (see inset).
    The initial set of parameters and the corresponding eigenvalues are
    represented by filled starting points. The arrows indicate the direction of
    progression. The filled triangle marks the position of the exceptional
    point in the parameter space and the corresponding complex energy of the
    degenerate resonances.}
\end{figure}

As an example Fig.\ \ref{fig:hydrogen_resonances_circle}
shows a typical result obtained in a numerical calculation for the exceptional
point labeled 14 in Table \ref{tab:hydrogen_exceptional_points}. The squares and
the diamonds represent each of the eigenvalues at different field strengths.
In this example, using 20 steps on the circle in the parameter space has been
sufficient to obtain a clear signature of the branch point singularity. The
``radius'' of the circle according to Eq.\ \eqref{eq:parameter_circle} was
$\delta = 0.01$.

\section{Description of the resonance energies in the vicinity of
  exceptional points}
\label{sec:hydrogen_properties}

\subsection{Effective two-dimensional matrix}
\label{sec:hydrogen_properties_new_model}

The most common case of an exceptional point consists of two resonances
forming a square root branch point, whose effects on the spectrum appear
in a close vicinity of the critical parameter values. There, only the two
resonances involved in the branch point singularity are important if they
are sufficiently separated from all further resonances, which in general can
be realized since only a narrow region of the complex energy plane around two
almost degenerate eigenvalues must be taken into account. In this case,
one can restrict the discussion to the subspace spanned by the two relevant
eigenvectors close to the exceptional point, which leads
to effective two-dimensional matrix models. Indeed, two-dimensional models
with only one complex parameter similar to the model introduced in Sec.\
\ref{sec:ep_simple_model} yield a good description of the two complex
eigenvalues in the vicinity of exceptional points \cite{Hei99}, and they can
also provide good results for the resonances of the hydrogen atom in external
fields. However, for some effects the actual matrix structure, which includes
the two real field strengths $\gamma$ and $f$, has to be taken into account.
We introduce a model adequate to describe the physical crossed-fields system.
It is given by a matrix, whose elements have the form
\begin{equation}
  M_{ij} = a^{(0)}_{ij} + a^{(\gamma)}_{ij} (\gamma-\gamma_0) + a^{(f)}_{ij} (f-f_0)
  \;, \quad i,j \in \{1,2\}
  \; . \label{eq:hydrogen_modelsmatrix}
\end{equation}
This two-dimensional matrix is still a very simple model because it includes
only the two states merging at the exceptional point and ignores couplings
to other levels. Furthermore, a linear dependence of the matrix elements on
the two field strengths is assumed, which is certainly only true for small
distances to the center point $(\gamma_0,f_0)$. In contrast with the simple
model used in Sec.\ \ref{sec:ep_simple_model} it includes two real parameters
with complex prefactors $a^{(0)}_{ij}$, $a^{(\gamma)}_{ij}$, $a^{(f)}_{ij}$ and
correctly reproduces the matrix shape of the full Hamiltonian to lowest order.
In a power series expansion its eigenvalues fulfill the relations
\begin{subequations}
  \begin{align}
    \lambda_1 + \lambda_2 &= \mathrm{tr}(\bm{M}) =
    A + B (\gamma-\gamma_0) + C (f-f_0) \; ,
    \label{eq:hydrogen_modelrelation1}  \\
    (\lambda_1 - \lambda_2)^2 &= \mathrm{tr}(\bm{M})^2 - 4 \mathrm{det}(\bm{M})
    = D + E (\gamma-\gamma_0) \notag \\ &\quad + F (f-f_0) 
    + G (\gamma-\gamma_0)^2 \notag \\ 
    &\quad + H (\gamma-\gamma_0)(f-f_0) + I (f-f_0)^2  
    \label{eq:hydrogen_modelrelation2}
  \end{align}
\end{subequations}
with new coefficients $A$, $B$, $C$, $D$, $E$, $F$, $G$, $H$, and $I$. Since
the eigenvalues do not change under a similarity transformation of the matrix
$\bm{M}$ and the explicit choice of the matrix is not relevant, the
representation  \eqref{eq:hydrogen_modelrelation1} and
\eqref{eq:hydrogen_modelrelation2} is more suitable than Eq.\
\eqref{eq:hydrogen_modelsmatrix}, in which more coefficients appear. The
coefficients can be determined by a fit of the exact quantum energies to Eqs.\
\eqref{eq:hydrogen_modelrelation1} and \eqref{eq:hydrogen_modelrelation2} in a
region of the parameter space in which only two resonances are relevant. A fit
for 6 different parameter sets yields the 9 coefficients $A$ to $I$ and, thus,
determines completely the two eigenvalues of the model
\eqref{eq:hydrogen_modelsmatrix} in dependence of the deviations from the
center point $(\gamma_0,f_0)$. Six differences $\lambda_1-\lambda_2$ are
required to determine $D$, $E$, $F$, $G$, $H$, and $I$. Three of the same
parameter sets can be used to determine $A$, $B$, and $C$ from the sum
$\lambda_1 +\lambda_2$.

\subsection{Shapes of the eigenvalue loops}
\label{sec:hydrogen_shapes}

\begin{figure}[tb]
  \centering
  \includegraphics[width=\columnwidth]{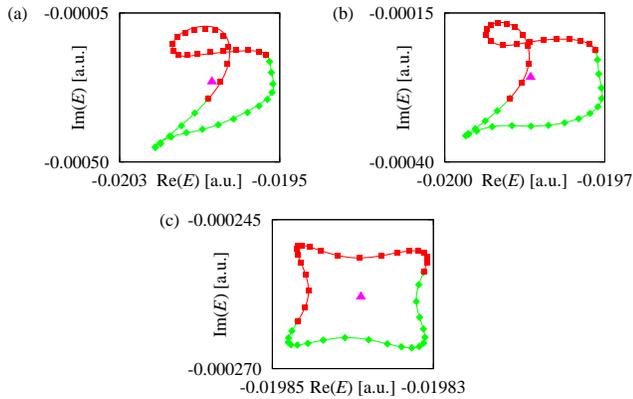}
  \caption{\label{fig:hydrogen_circles_radii}(Color online) Paths of the two
    eigenvalues considered in Fig.\ \ref{fig:hydrogen_resonances_circle}
    for different radii of the parameter space circle 
    \eqref{eq:parameter_circle}, where the center $(\gamma_0,f_0) = 
    (\gamma^\mathrm{(EP)},f^\mathrm{(EP)})$ was always chosen to be exactly the
    exceptional point. (a) $\delta=3\times 10^{-2}$, (b) $\delta = 10^{-2}$,
    (c) $\delta = 10^{-4}$. The position of the resonances at
    $(\gamma^\mathrm{(EP)},f^\mathrm{(EP)})$ is marked by a triangle in each
    figure. The squares and diamonds represent the exact quantum resonances.
    The lines represent the eigenvalues of the two-dimensional matrix model 
    \eqref{eq:hydrogen_modelsmatrix}, whose coefficients have been
    fitted to the numerical results of the exact quantum calculations.}
\end{figure}

As can be seen in Fig.\ \ref{fig:hydrogen_resonances_circle} the shape of
the paths covered by the two energy eigenvalues differs considerably from
the semicircle form obtained for a small loop in the simple two-dimensional
matrix model in Fig.\ \ref{fig:ep_most_simple_model}(b). A good description of
the complicated behavior is possible with the model
\eqref{eq:hydrogen_modelrelation1}, \eqref{eq:hydrogen_modelrelation2}. The
results of a more detailed investigation of the phenomenon with the matrix
model are shown in Fig.\ \ref{fig:hydrogen_circles_radii}.
A circle around the exceptional point, which is always located 
exactly at the center of each figure, is performed for three different radii.
For radii $\delta = 0.03$ and $\delta = 0.01$ (see Figs.\ 
\ref{fig:hydrogen_circles_radii}(a) and (b)) the complicated structure
already known from Fig.\ \ref{fig:hydrogen_resonances_circle}(a) appears.
The deformations of the eigenvalue paths can be reproduced with
the matrix \eqref{eq:hydrogen_modelsmatrix}. 
The lines in Figs.\ \ref{fig:hydrogen_circles_radii}(a) and (b) represent
the positions of the two model eigenvalues $\lambda_1$ and $\lambda_2$ for the
same parameter space circle which was used for the exact quantum resonances. The
very good agreement demonstrates that it is possible to describe the local
structure of the resonances at an exceptional point with a simple
two-dimensional model that ignores the influence of further resonances even if
complicated structures in the eigenvalue paths appear.

Fig.\ \ref{fig:hydrogen_circles_radii}(c) shows a circle around the same
exceptional point for the much smaller radius $\delta=10^{-4}$, where the shape
of the paths becomes more similar to the semicircle known from Fig.\
\ref{fig:ep_most_simple_model}(b). The exceptional point marked by the
triangle in Fig.\ \ref{fig:hydrogen_circles_radii}(c) now is located at the
center of the enclosing eigenvalue trajectories. However, the loop still is not
a perfect semicircle. This effect is a result of the dependence on two real
parameters with two complex prefactors as introduced in the model 
\eqref{eq:hydrogen_modelsmatrix}. While the models using one complex parameter
(cf.\ Sec.\ \ref{sec:ep_simple_model}) lead to a perfect semicircle, as was
demonstrated with the power series expansion
\eqref{eq:most_simple_model_expansion}, this is not the case for the description
\eqref{eq:hydrogen_modelsmatrix}. A short calculation shows that a fractional
power series expansion similar to Eq.\ 
\eqref{eq:most_simple_model_expansion} for the model
\eqref{eq:hydrogen_modelrelation1}, \eqref{eq:hydrogen_modelrelation2} 
to lowest order reads
\begin{equation}
  \lambda_{1,2} = \begin{cases}
    \frac{A}{2} \pm \frac{1}{2} \sqrt{ \frac{U}{2}
      \delta} \; \sqrt{1+\frac{V}{U} \mathrm{e}^{-\mathrm{i} 2 \varphi}}
    \; \mathrm{e}^{\mathrm{i} \varphi/2} \; , & \text{if } |V|<|U| \\
    \frac{A}{2} \pm \frac{1}{2} \sqrt{\frac{V}{2}
      \delta} \; \sqrt{1+\frac{U}{V} \mathrm{e}^{\mathrm{i} 2 \varphi}}
    \; \mathrm{e}^{-\mathrm{i} \varphi/2} \; , & \text{if } |V|>|U|
  \end{cases}
  \label{eq:hydrogen_properties_model_lambda_expansion}
\end{equation}
with $U = E\gamma_0 + F f_0$, and $V = E\gamma_0 - F f_0$. The second
term under the second square root is large enough to have a considerable
influence and leads to the modulation of the ``radius'' during the traversal
of the semicircle evident in Fig.\ \ref{fig:hydrogen_circles_radii}(c). Again,
the model fitted to the numerical data, and rendered by the continuous lines
in the figure, perfectly reproduces the exact behavior.

\begin{figure}[tb]
  \centering
  \includegraphics[width=\columnwidth]{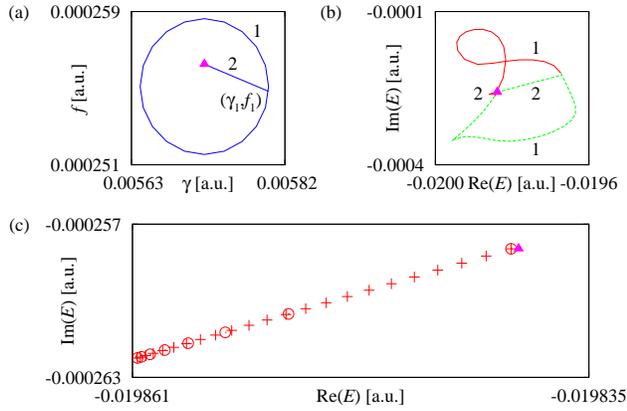}
  \caption{\label{fig:hydrogen_intersection}(Color online) (a) A circle in the
    parameter space (labeled 1) is chosen such that it passes through the
    parameter set $(\gamma_1,f_1)$ at which one of the energy eigenvalues is
    identical with its position at the exceptional point. A straight line
    (labeled 2) connects the parameters of the exceptional point and the
    parameter set $(\gamma_1,f_1)$.
    (b) Paths of the two numerically exact complex energies for the circle 
    and the straight line described in (a). (c) Magnification of the path
    of the eigenvalue denoted by the red solid line in (b) for the parameter
    line labeled 2 in (a). The eigenvalue departs from the point of degeneracy
    (circles) and returns to its initial value (plus symbols).}
\end{figure}

Intersections of an eigenvalue path with itself are observed, which means
that the eigenvalue can have the same complex energy for two different
parameter sets. In Fig.\ \ref{fig:hydrogen_circles_radii}(a) one can even see
that the position of the exceptional point lies outside the area enclosed by
the two eigenvalue paths. This is possible if \emph{one} of the two eigenvalues
(in this example obviously the resonance denoted by red squares) has
the same complex energy as the one at the exceptional point for a second
parameter set. The crossing of an eigenvalue path with the position of an
exceptional point in the complex energy plane is shown in Fig.\
\ref{fig:hydrogen_intersection}.
The parameter space circle labeled 1 in Fig.\ \ref{fig:hydrogen_intersection}(a)
is chosen such that it passes directly through the point $(\gamma_1,f_1)$ at
which one of the resonances returns to its initial position at the exceptional
point as can be seen in Fig.\ \ref{fig:hydrogen_intersection}(b).
Additionally, the straight line labeled 2 in Fig.\
\ref{fig:hydrogen_intersection}(a) is traversed. For this line one observes 
that both resonances leave the position of the exceptional point. The resonance
denoted by the solid red line returns to its original position on the same
path as can be seen in the magnified illustration in Fig.\
\ref{fig:hydrogen_intersection}(c).

\section{Dipole matrix elements and the photoionization cross
  section at exceptional points}
\label{sec:hydrogen_dip_behavior}

In the correct inner product, which has to be used for the complex rotated
states, no complex conjugation of the factors $\mathrm{e}^{\mathrm{i} \theta}$
(cf.\ Eq.\ \eqref{eq:crot_r}) may be performed \cite{Moi98}. Only the intrinsic
complex parts of the wave functions are complex conjugated. An important
consequence is that resonance wave functions normalized with respect to this
inner product diverge at the branch points, which is typical of exceptional
points \cite{Hei99}. 
In the simple two-dimensional matrix model introduced in Sec.\
\ref{sec:ep_simple_model} that behavior can be observed directly for the
normalized eigenvectors 
\begin{equation}
  \bm{x}_{1,2}(\kappa) = \frac{1}{\sqrt{\kappa^2 + \left ( 1 \mp
        \sqrt{1+\kappa^2}\right )^2}}
  \left ( \begin{array}{c} 
      -\kappa \\ 1\mp\sqrt{1+\kappa^2} \end{array} \right ) 
  \label{eq:ep_normalized_vector}
\end{equation} 
at the exceptional points $\kappa = \pm \mathrm{i}$.

One may wonder whether or not the diverging behavior carries over to measurable
physical quantities. In particular, the photoionization cross section is
important for the observation of  resonances in experiments. For example, in
Ref.\ \cite{Car07b} it was proposed to measure the photoionization cross section
for parameter sets located on a closed curve and to extract the complex
energies of the resonances from the cross section with the harmonic inversion
method. This procedure allows for searching the permutation behavior in
experimental data. 

As presented by Rescigno and McKoy \cite{Res75} the dipole matrix elements
\begin{equation}
  P_j(E) = \langle \Psi_0 | D(\bm{r}) | \Psi_j(\theta) \rangle
  \label{eq:hydrogen_dip_elements}
\end{equation} 
and the photoionization cross section
\begin{equation}
  \sigma(E) = 4\pi \alpha_\mathrm{elm} (E-E_0) \mathrm{Im} 
  \left ( \sum_j \frac{\left < \Psi_0 | D | \Psi_j(\theta) \right >^2}{E_j - E} 
  \right ) \; ,
  \label{eq:hydrogen_form_cross_section}
\end{equation}
can straightforwardly be calculated once the energies obtained from the
complex rotated Hamiltonian and the corresponding eigenvectors are at hand.
In Eq.\ \eqref{eq:hydrogen_form_cross_section} $D$ is the dipole operator
in atomic units for a given direction of polarization and $\alpha_\mathrm{elm}$
is the fine-structure constant. The bound state, whose energy $E_0$ is supposed
to be known, is represented by $\Psi_0$. The ionized states are labeled
$\Psi_j$, where the rotated eigenfunctions calculated by the complex rotation
method are used. In converged spectra, the result is independent of $\theta$.

Due to the dependence of Eqs.\ \eqref{eq:hydrogen_dip_elements} and
\eqref{eq:hydrogen_form_cross_section} on the resonance wave functions
$\Psi_j$ a remarkable behavior at exceptional points occurs. The diverging
behavior of the wave functions must carry over to the dipole matrix elements
and, indeed, the numerical results show that the resonance wave functions
obtained by the complex rotation method lead to diverging dipole matrix
elements. This is, however, not an observable physical property because the
\emph{single} dipole matrix elements of the two identical wave functions at an
exceptional point are not accessible. In particular, the photoionization cross
section behaves regularly and does not diverge at an exceptional point as will
be shown below.

Again, a two-dimensional matrix model helps in understanding these effects. To
keep the discussion transparent, the simplest example, namely the symmetric
matrix model introduced in Sec.\ \ref{sec:ep_simple_model} is used, 
the results, however, are valid for all complex symmetric matrices.
Note that in particular there is no difference visible in the behavior of the
eigenvalues between the lowest-order power series expansions
\eqref{eq:most_simple_model_expansion} and
\eqref{eq:hydrogen_properties_model_lambda_expansion} if only the distance
$\delta$ (or $\varrho$) is varied and the angle $\varphi$ is kept constant.
The normalized eigenvectors, which correspond to the resonance wave
functions $\Psi_j$, in this model are given by Eq.\ 
\eqref{eq:ep_normalized_vector}. The form of the dipole matrix elements
\eqref{eq:hydrogen_dip_elements} is represented by a product $P_{1,2} =
\bm{y}\cdot \bm{x}_{1,2}$ of the eigenvectors $\bm{x}_{1,2}$ with an arbitrary
vector $\bm{y} = (y_1,y_2)$. If the squares of the dipole matrix elements,
which are required for the photoionization cross section, are calculated for a
small complex deviation $\delta$ from one of the two exceptional points,
$\kappa = \mathrm{i} + \delta$, a fractional power series expansion shows that
the single contributions
\begin{multline}
  \bar{P}_{1,2}^2 = (\bm{y}\cdot \bm{x}_{1,2})^2 =
  \pm \frac{\mathrm{e}^{\mathrm{i} 3\pi/4}}{2\sqrt{2}} 
  \frac{(y_1 + \mathrm{i} y_2)^2}{\sqrt{\delta}} + \frac{1}{2}(y_1^2+y_2^2) \\ 
  \pm \frac{\mathrm{e}^{\mathrm{i} 5\pi/4}}{8\sqrt{2}}
  (y_1^2 - 6\mathrm{i} y_1 y_2 - y_2^2) \sqrt{\delta} + \mathrm{O}(\delta)
  \label{eq:hydrogen_dip2_div}
\end{multline}
diverge as $1/\sqrt{\delta}$. It is interesting to note that the sum
of both contributions always has the \emph{exact} value
\begin{equation}
  (\bm{y}\cdot \bm{x}_1)^2+(\bm{y}\cdot \bm{x}_2)^2 =  y_1^2 + y_2^2 \; ,
  \label{eq_hydrogen_dip2_sum}
\end{equation}
independent of the parameter $\kappa$, i.e., of the presence of the
exceptional point, and of the matrix used. What is more interesting, however, 
is the sum
\begin{equation}
  \bar{\sigma} = \frac{(\bm{y}\cdot \bm{x}_1)^2}{\lambda_1 - E} + 
  \frac{(\bm{y}\cdot \bm{x}_2)^2}{\lambda_2 - E} \; ,
\end{equation}
which describes the contribution of the two resonances to the photoionization
cross section with the eigenvalues \eqref{eq_most_simple_ev} and a real
variable $E$ representing the energy. Here, one can also look at the
contributions of the single eigenvalues, and obtains in a fractional power
series expansion around the branch point
\begin{multline}
  \bar{\sigma}_{1,2} = \frac{(\bm{y}\cdot \bm{x}_{1,2})^2}{\lambda_{1,2} - E} =
  \pm \frac{\mathrm{e}^{\mathrm{i} 7\pi/4}}{2\sqrt{2}}
  \frac{(y_1 + \mathrm{i} y_2)^2} {E \sqrt{\delta}} + f_1(E,\bm{y}) \\
  \pm f_2(E,\bm{y}) \sqrt{\delta} + f_3(E,\bm{y}) \delta 
  + \mathrm{O}(\delta^{3/2})
\end{multline}
with rather complicated expressions $f_1(E,\bm{y})$, $f_2(E,\bm{y})$, and 
$f_3(E,\bm{y})$ which do not depend on $\delta$. These parts diverge, however,
$\bar{\sigma}_1$ and $\bar{\sigma}_2$ alone are not observable. The sum
contributes to the photoionization cross section, and, in particular, at
the exceptional point ($\delta = 0$) the two resonances overlap. For the sum
one finds
\begin{equation}
  \bar{\sigma} = 2f_1(E,\bm{y}) + 2f_3(E,\bm{y}) \delta + \mathrm{O}(\delta^2)
  \; , \label{eq:hydrogen_cs_linear}
\end{equation}
that is, the photoionization cross section converges linearly to a constant
value at the branch point.

Numerical calculations for the hydrogen spectra discussed in this paper
\begin{figure}[tb]
  \centering
  \includegraphics[width=\columnwidth]{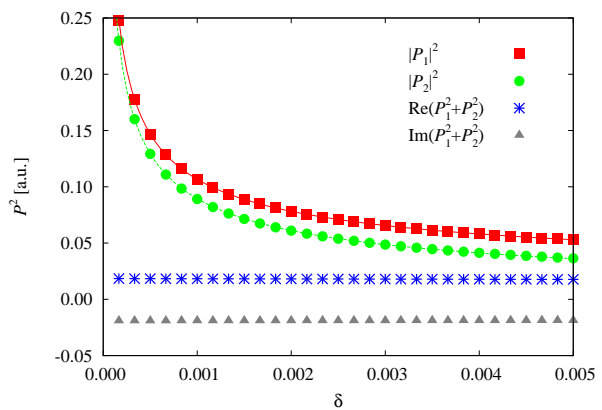}
  \caption{\label{fig:hydrogen_dipole_div}(Color online) Squares $P^2$ of the
    dipole matrix elements between a wave function corresponding to a bound
    state and the two states evolving from the exceptional point labeled 12
    in Table \ref{tab:hydrogen_exceptional_points} at field strengths according
    to the form \eqref{eq:parameter_circle} with the constant angle $\varphi 
    = 0.7$ and $(\gamma_0,f_0) = (\gamma^\mathrm{(EP)}, f^\mathrm{(EP)})$, 
    $\delta$ is the distance from the critical values of the exceptional point.
    For both states the square modulus diverges in the form of a reciprocal
    square root of the distance. A fit of the data points to a function of the
    form $a/\sqrt{\delta} + b$, which is expected from Eq.\
    \eqref{eq:hydrogen_dip2_div}, is shown by the solid red and dashed green
    lines. The real and the imaginary part of the sum $P_1^2+P_2^2$ is also
    drawn.}
\end{figure}
demonstrate the applicability of the simple model. For this purpose, the square
modulus of the dipole matrix elements \eqref{eq:hydrogen_dip_elements} and the
photoionization cross section \eqref{eq:hydrogen_form_cross_section} are
calculated on a straight line of the form \eqref{eq:parameter_circle} for a
constant angle $\varphi$ and variable distance $\delta$ from the exceptional
point $(\gamma_0,f_0) = (\gamma^\mathrm{(EP)}, f^\mathrm{(EP)})$. The results
have been verified for different angles $\varphi$.
Fig.\ \ref{fig:hydrogen_dipole_div} shows the squares of the two
isolated dipole matrix elements for the resonances which form the exceptional
point labeled 12 in Table \ref{tab:hydrogen_exceptional_points}.
Both matrix elements behave as predicted by the approximation
\eqref{eq:hydrogen_dip2_div} of the simple two-dimensional model. The single
terms $|P_1|^2$ and $|P_2|^2$ diverge in the form of a reciprocal square root
of $\delta$ which is shown by a fit of the numerical results to the function
\begin{equation}
  P_\mathrm{fit}(\delta) = \frac{a}{\sqrt{\delta}} + b \; .
\end{equation}
An excellent agreement can be observed. Additionally, the real and imaginary
parts of the sum $P_1^2+P_2^2$ are plotted. As expected from
Eq.\ \eqref{eq_hydrogen_dip2_sum} for the two-dimensional model, this sum
has a constant value.

\begin{figure}[tb]
  \centering
  \includegraphics[width=\columnwidth]{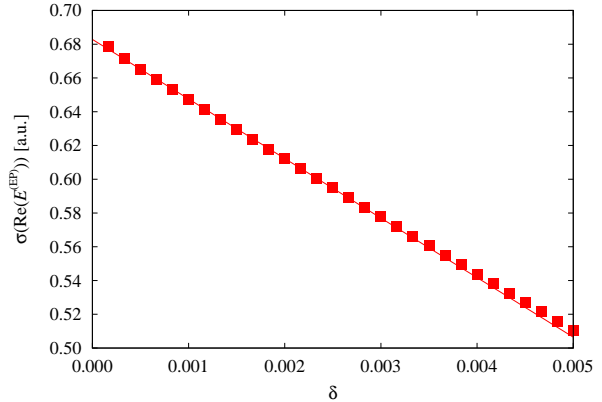}
  \caption{\label{fig:hydrogen_cross_section_div}(Color online) Photoionization
    cross section $\sigma(\mathrm{Re}(E^\mathrm{(EP)}))$ with the bound state
    energy $E_0=-0.125$ ($n=2$, p-orbital) as a function of the distance
    $\delta$ from the critical field values of the exceptional point. The
    numerical data (dots) converge linearly to a constant value as is expected
    from Eq.\ \eqref{eq:hydrogen_cs_linear}. Again, the line
    \eqref{eq:parameter_circle} with constant angle $\varphi = 0.7$ and
    $(\gamma_0,f_0) = (\gamma^\mathrm{(EP)},f^\mathrm{(EP)})$ is used.}
\end{figure}

The photoionization cross section $\sigma(\mathrm{Re}(E^\mathrm{(EP)}))$
evaluated at the real part of the energy $E^\mathrm{(EP)}$ at the exceptional
point according to Eq.\ \eqref{eq:hydrogen_form_cross_section} is shown in
Fig.\ \ref{fig:hydrogen_cross_section_div} as a function of the distance
parameter $\delta$.
As can be seen directly in the figure the numerical data points (red points)
converge linearly to a constant value for $\delta \to 0$. The red line
represents a fit to the function
\begin{equation}
  \sigma_\mathrm{fit}(\mathrm{Re}(E^\mathrm{(EP)})) = a + b \delta
\end{equation}
whose form is expected from the power series expansion
\eqref{eq:hydrogen_cs_linear} of the two-dimensional model. The comparison
shows an excellent agreement.

\section{Structures with three resonances}
\label{sec:hydrogen_properties_three}

Beyond the typical square root branch point behavior studied in most physical
examples, higher branch points connecting more than two eigenvalues are
possible. In particular, the possibility of a cubic root branch point in a
three-dimensional symmetric matrix is the topic of current studies
\cite{Hei08}. For a complex symmetric matrix a coalescence of $N$ levels
requires $(N^2+N-2)/2$ real parameters \cite{Hei08}. Thus, for $N=3$ five
parameters are necessary. For the hydrogen atom this means that
additionally to the two field strengths $\gamma$ and $f$ three further
parameters have to be introduced. Locating exceptional points in a
five-dimensional parameter space is expected to be a laborious task
considering the difficulties encountered already for two parameters.
However, as will be shown below, a combination of three resonances
strongly related to a cubic root branch point is observable in spectra
of the crossed-fields hydrogen atom. In particular the permutation
behavior of a cubic root branch point, where three resonances are permuted
and three circles in the parameter space are required to restore the
original situation, can be found.

\begin{figure}[tb]
  \centering
  \includegraphics[width=\columnwidth]{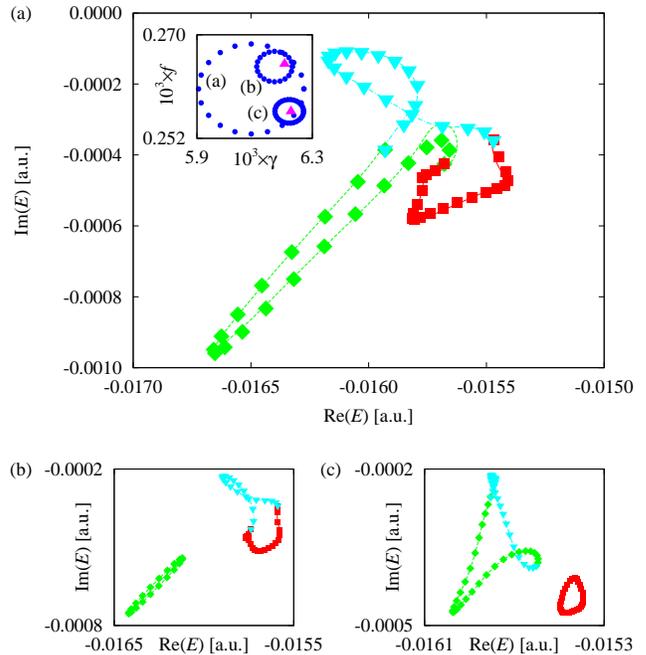}
  \caption{\label{fig:hydrogen_three_resonances}(Color online) Illustration of
    a structure in the vicinity of two exceptional points in which three
    resonances are involved. Each of the resonances is denoted by a different
    symbol and color. The inset of Fig.\ (a) shows the three parameter space
    loops (circles) and the position of the two exceptional points (up
    triangles). 
    In Fig.\ (a) the parameter space loop (indicated by dots) is chosen such
    that both exceptional points are located within the circular area. The
    resonance marked by down triangles forms a branch point singularity with two
    further states at two different parameter values, which is shown by two
    additional parameter space loops (small circles in the inset) in whose
    circular area only one of the two exceptional points is located
    (Figs.\ (b) and (c)). The solid lines mark the eigenvalues of the
    three-dimensional model 
    \eqref{eq:hydrogen_3dmodel_a}-\eqref{eq:hydrogen_3dmodel_c} fitted to the
    numerical data. }
\end{figure}

An example where three resonances come into play is given in Fig.\
\ref{fig:hydrogen_three_resonances}.
In Fig.\ \ref{fig:hydrogen_three_resonances}(a) obviously a 
permutation of three eigenvalues indicated by points with three different 
symbols and colors can be observed for a closed loop in the parameter space.
From that behavior one might assume that indeed an exceptional point
consisting of three resonances, which form a cubic root branch point
singularity, was detected. A more detailed analysis shows, however, that this is
not the case. But there is a close relationship with a triple coalescence.
There are two exceptional points located in the circular area of the parameter
space circle of the type \eqref{eq:parameter_circle} used in Fig.\ 
\ref{fig:hydrogen_three_resonances}(a). This can directly be shown if one
chooses two different parameter loops with two different center points
$(\gamma_0,f_0)$ and smaller radii $\delta$, which is done in Figs.\ 
\ref{fig:hydrogen_three_resonances}(b) and (c). Then one observes that
there are two exceptional points at which two of the three resonances form a
square root branch point. The resonance denoted by down triangles in 
Fig.\ \ref{fig:hydrogen_three_resonances} is involved in both exceptional
points. For the large parameter space circle used for Fig.\
\ref{fig:hydrogen_three_resonances}(a) the two different exceptional points
cannot be resolved and the eigenvalue paths form the permutation of three
resonances which looks like a triple coalescence in the form of a cubic
root branch point. Although no exact triple coalescence was found the finding
of the situation depicted here is already a remarkable result. Two exceptional
points located close to each other in the parameter space are required.
Furthermore, only three resonances may be connected with these exceptional
points, i.e., one of them must be connected with both branch points. If three 
additional parameters were available, a coalescence of both exceptional points
would be possible or, in other words, the additional parameters would be
necessary to shift the three resonances such that they form a cubic root branch
point in the five-dimensional parameter space.

Of course, it is not possible to describe a behavior of this kind with the
two-dimensional matrix models used so far. At least a three-dimensional model
is required to simulate three energy eigenvalues connected with each other.
Indeed, it can be shown that it is possible to reconstruct the structures shown
in Fig.\ \ref{fig:hydrogen_three_resonances} by a three-dimensional matrix
model. Similar to the ansatz \eqref{eq:hydrogen_modelsmatrix} we expand the
matrix elements in a power series in the two field strengths $\gamma$ and $f$
around a center point, and specifically assume the matrix to be symmetric. To
model the behavior of its eigenvalues $\lambda$, we fit the coefficients
of the characteristic polynomial
\begin{subequations}
  \begin{equation}
    \lambda^3 + a \lambda^2 + b \lambda + c = 0 
    \label{eq:hydrogen_3dmodel_polynomial}
  \end{equation}
  to the exact numerical results. Note that one has the familiar relations
  \begin{align}
    a &= - (\lambda_1 + \lambda_2 + \lambda_3) \; ,
    \label{eq:hydrogen_3dmodel_a} \\
    b &= \lambda_1 \lambda_2 + \lambda_1 \lambda_3 + \lambda_2 \lambda_3 \; ,\\
    c &= - \lambda_1 \lambda_2 \lambda_3 \; . \label{eq:hydrogen_3dmodel_c} 
  \end{align}
\end{subequations}
For the discussion in Fig.\ \ref{fig:hydrogen_three_resonances} a power series
expansion of the coefficients $a$, $b$, and $c$ up to third order in both field
strengths was included, which leads to 30 terms for all three coefficients,
i.e., 10 combinations of field strengths are required to obtain 10 triples of
eigenvalues for the relations 
\eqref{eq:hydrogen_3dmodel_a}--\eqref{eq:hydrogen_3dmodel_c}. The
eigenvalues in the three-dimensional matrix model are shown as solid lines in
Figs.\ \ref{fig:hydrogen_three_resonances}(a), (b), and (c) and agree very
well with the numerically exact resonances. This shows that it is sufficient to
only take the three resonances into account to explain their behavior. The
influence of further resonances can be ignored in the investigation of the
threefold permutation. The model can even be used to predict the positions of
the two exceptional points located within the parameter space loop. For the
case shown in Fig.\ \ref{fig:hydrogen_three_resonances}(a) the model predicts
the positions of the two exceptional points labeled 15 and 16
in Table \ref{tab:hydrogen_exceptional_points} at
$(\gamma_1 = 6.12\times 10^{-3}, f_1 = 2.53\times 10^{-4})$ and  
$(\gamma_2 = 6.15\times 10^{-3}, f_2 = 2.68\times 10^{-4})$, respectively, which
well approximates the results of the full quantum treatment.

\section{Connection with avoided level crossings}
\label{sec:hydrogen_avoided}

There is a close relation between avoided level crossings of the real
energies of bound states and exceptional points. As has been demonstrated,
the level repulsions of bound states of a Hermitian Hamiltonian which depends
on one real parameter are associated with an exceptional point if the parameter
is continued into the complex plane \cite{Hei90,Hei99}. A similar effect
appears with resonances of open quantum systems. Here one can observe
crossings or avoided crossings of either the positions or the widths of the
resonances for lines in the parameter space which do not run over the
exceptional point. This behavior has, e.g., been discussed for the resonances
in microwave cavities \cite{Dem01}.

\begin{figure}[tb]
  \centering
  \includegraphics[width=\columnwidth]{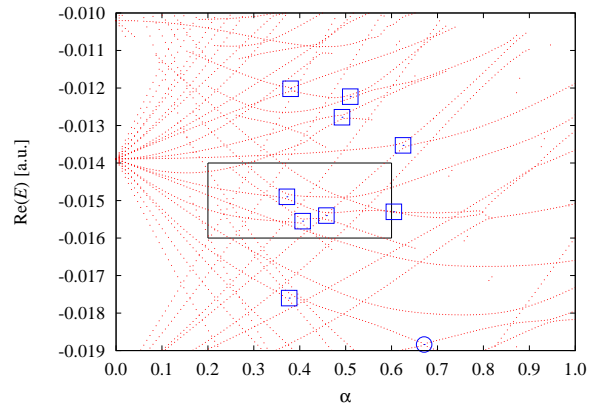}
  \caption{\label{fig:hydrogen_avoided}(Color online) Real part of the complex
    energy as a function of the one-dimensional parameter $\alpha$, $\gamma =
    0.80516 \times 10^{-2} \alpha$, $f = 0.32169 \times 10^{-3} \alpha$, defined
    in Eqs.\ \eqref{eq:hydrogen_1dline_gamma} and
    \eqref{eq:hydrogen_1dline_f}. The blue circle marks an exceptional point.
    Avoided crossings (marked by blue squares) appear, which are related to
    exceptional points. Only resonances with $|\mathrm{Im}(E)| < 0.0005$ are
    drawn to not overload the figure. The black frame marks the region
    investigated in detail in Fig.\ \ref{fig:hydrogen_avoided_detail}(a) to
    demonstrate how an avoided crossing leads to an exceptional point if the
    parameters are varied.}
\end{figure}

Avoided level crossings of the energies or widths can also be observed in
spectra of the hydrogen atom in external fields. Fig.\
\ref{fig:hydrogen_avoided} shows the real part of the complex energy as a
function of a parameter $\alpha$ which defines a straight line of the form
\begin{subequations}
  \begin{align}
    \gamma &= 0.80516 \times 10^{-2} \alpha \label{eq:hydrogen_1dline_gamma}
    \; , \\
    f &= 0.32169 \times 10^{-3} \alpha \label{eq:hydrogen_1dline_f} \; , \\
    \gamma/f &= 25.03
  \end{align}
\end{subequations}
in the $(\gamma,f)$-space.
The line is chosen such that it passes directly through the exceptional point
labeled 12 in Table \ref{tab:hydrogen_exceptional_points}. The
position of the exceptional point is marked by the blue circle in the figure.
In the vicinity of the parameter space line further exceptional points are
located. Since the line does not pass through these critical parameter values, 
the exceptional points only manifest themselves by avoided crossings in the
energies or the widths. The avoided energy (real part) crossings are shown in
the figure. They are marked by blue squares. The close connection between
avoided crossings and branch point singularities of exceptional points can be
shown very directly. In all cases marked in Fig.\ \ref{fig:hydrogen_avoided} 
it is possible to vary the parameters $\gamma$ and $f$ such
that a coalescence of the two eigenvalues forming the avoided crossing
is achieved. That is, all avoided crossings shown in the diagram are
associated with exceptional points of the corresponding resonances.
In all cases where the description of the energy range under consideration 
is possible with the model \eqref{eq:hydrogen_modelsmatrix}, i.e., always
if only two resonances are involved, the adjustment of the parameters
$\gamma$ and $f$ leads to a branch point. In other words, exceptional points
are indeed found to be the origin of narrow avoided level crossings.

\begin{figure}[tb]
  \centering
  \includegraphics[width=\columnwidth]{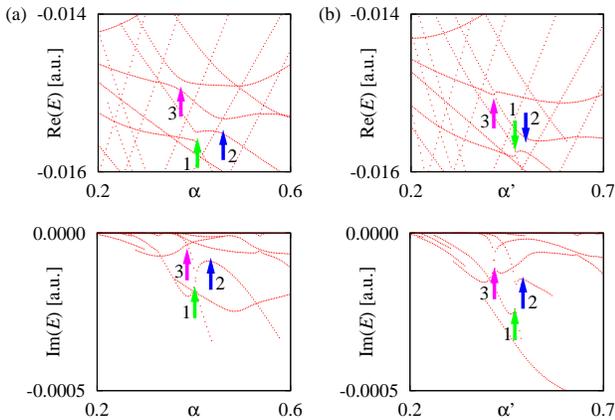}
  \caption{\label{fig:hydrogen_avoided_detail}(Color online) (a) Detailed
    illustration of the region marked by the black frame in Fig.\
    \ref{fig:hydrogen_avoided}.
    Three avoided energy crossings are present. In addition, the
    imaginary part is shown, which exhibits two crossings (arrows 1 and 3) and
    one avoided crossing (arrow 2). (b) The same region
    is shown for the field strengths varied on the line $\gamma = 0.65378
    \times 10^{-2} \alpha'$, $f = 0.28213 \times 10^{-3} \alpha'$ given by
    Eqs.\ \eqref{eq:hydrogen_1dline2_gamma} and 
    \eqref{eq:hydrogen_1dline2_f}.
    One of the three exceptional points is located on this line and the
    corresponding avoided energy crossing marked by arrow 3 has been
    transformed into a branch point singularity, where both the real and
    imaginary parts are identical. The other two encounters show at least one
    avoided crossing in the real or in the imaginary part.}
\end{figure}

A more detailed illustration is given in Fig.\
\ref{fig:hydrogen_avoided_detail}.
The region marked by the black frame in Fig.\ \ref{fig:hydrogen_avoided} is
magnified in Fig.\ \ref{fig:hydrogen_avoided_detail}(a). Three of the 
avoided level crossings are included and, in addition, the imaginary
part of the resonances is shown. The encounter of two resonances marked
by the arrow labeled 1 shows an avoided crossing in the real part and a
crossing in the imaginary part. The arrow 2 exhibits avoided crossings
in the real as well as in the imaginary part. The encounter marked by
arrow 3 shows an avoided crossing of the real part which is
accompanied by a crossing in the imaginary part. Its behavior changes on a
neighboring line in the parameter space defined by
\begin{subequations}
  \begin{align}
    \gamma &= 0.65378 \times 10^{-2} \alpha' \label{eq:hydrogen_1dline2_gamma}
    \; , \\
    f &= 0.28213 \times 10^{-3} \alpha' \; , \label{eq:hydrogen_1dline2_f} \\
    \gamma/f &= 23.17 \; .
  \end{align}
\end{subequations}
Now the avoided energy crossing is changed into a crossing, which
can be seen in Fig.\ \ref{fig:hydrogen_avoided_detail}(b). The line
\eqref{eq:hydrogen_1dline2_gamma}, \eqref{eq:hydrogen_1dline2_f} runs
exactly through the corresponding exceptional point. As a consequence
the avoided crossing is transformed into a branch point degeneracy at which
both the real and imaginary parts are identical, whereas the other two
encounters do not form a degeneracy. They belong to different exceptional
points which are not hit by the line. The encounter labeled by arrow 2
forms, again, an avoided crossing both in the real and in the imaginary part.
The third encounter, which is marked by arrow 1 has inverted its
behavior. Now, it forms a real part crossing and an avoided crossing of the
imaginary part.

The connection between avoided crossings and exceptional points provides
an additional possibility of detecting the branch point singularities in 
spectra of the hydrogen atom in external fields. Exceptional points can be
found by plotting the real part of the complex resonance energies as a function
of \emph{one} parameter, similar to $\alpha$ in Eqs.\ 
\eqref{eq:hydrogen_1dline_gamma} and \eqref{eq:hydrogen_1dline_f}.
If avoided crossings of the energy are found, the identification of the
exceptional point can be performed by the method presented in Sec.\
\ref{sec:hydrogen_search}.

\section{Conclusion and outlook}
\label{sec:conclusion}

Exceptional points are a feature that can emerge in parameter-dependent open
quantum systems with decaying unbound states. They are branch point
singularities at which two eigenstates of a non-Hermitian Hamiltonian coalesce.
The topic of this paper was to investigate these exceptional points in the
spectra of atoms in external fields. It has been shown, that the branch point
singularities can be found by the permutation of two eigenvalues when an
exceptional point is encircled in the parameter space. This method works
reliably.

The effects of exceptional points on spectra of the hydrogen atom in external
fields have been analyzed in detail. In a close region around the branch
points, it is possible to describe the two branching eigenstates by
$2\times 2$-matrix models and to explain the structure of the loops the
eigenvalues traverse for closed paths in the parameter space. 
The study of dipole matrix elements in the vicinity of branch points has
revealed remarkable properties. While single dipole matrix elements diverge in
the presence of exceptional points this is not the case for observable
physical quantities such as the photoionization cross section. Both behaviors
can be explained by a simple matrix model, which provides an excellent
qualitative description of the properties in the local vicinity of the branch
point singularity. 

Beyond the typical square root branch points the spectra of the hydrogen atom
exhibit structures in which three resonances are connected via two exceptional
points. One of the three resonances is involved in both branch points. This
finding is especially important because it is closely related to a cubic root
branch point, at which all three resonances coalesce. The direct adjustment of
such a cubic root branch point would require five external parameters and
therefore is not possible in the system investigated here.  
Furthermore, the close relation of avoided level crossings and exceptional
points could be confirmed in the resonances of the hydrogen atom in external
fields.

As an outlook, it should be possible to extend the two-dimensional matrix 
model introduced in this paper to develop a further, and possibly very fast,
method to determine the exact position of an exceptional point. Once its
existence has been detected by the permutation of two eigenvalues the fit
to the two-dimensional matrix model can be used to predict the position of the
degeneracy in parameter space by solving Eq.\ \eqref{eq:hydrogen_modelrelation2}
for $\lambda_1 = \lambda_2$ which is a very easy task. The improved position
of the exceptional point can be used to perform a smaller parameter space
circle and to apply the procedure iteratively.

With the procedure proposed in Ref.\ \cite{Car07b} it should be possible to
verify experimentally the effects of exceptional points on the spectra of atoms
in external fields discussed in this paper. In particular, the permutation
behavior to prove their existence, the influence on the photoionization cross
section, and the close relation to avoided level crossings should be observable.

\begin{acknowledgments}
  This work was supported by Deutsche Forschungsgemeinschaft. H.C. 
  is grateful for support from the Landesgraduiertenf\"orderung of
  the Land Baden-W\"urttemberg.
\end{acknowledgments}

\end{document}